\documentclass[conference, 10pt]{IEEEtran}
\IEEEoverridecommandlockouts
\usepackage[dvipsnames]{xcolor}
\usepackage{subcaption}
\usepackage{graphicx}
\usepackage{textcomp}
\usepackage{xcolor}
\usepackage{svg}
\usepackage{float}
\usepackage{amsmath}
\usepackage{multicol}
\usepackage{array}
\usepackage{mwe}
\usepackage[noend]{algpseudocode}
\usepackage[ruled, lined, longend]{algorithm2e}
\usepackage{multirow}
\usepackage{arydshln}
\usepackage{adjustbox}
\usepackage[thinc]{esdiff}
\usepackage{booktabs} 
\usepackage{pifont}
\def\BibTeX{{\rm B\kern-.05em{\sc i\kern-.025em b}\kern-.08em
    T\kern-.1667em\lower.7ex\hbox{E}\kern-.125emX}}

\usepackage{hyperref}

\usepackage{amsfonts}
\usepackage{mathabx}

\newif\ifsubmit
\submitfalse
\ifsubmit
  \newcommand{\todo}[1]{}
  \newcommand{\ye}[1]{}
  \newcommand{\zhiheng}[1]{}
  \newcommand{\yian}[1]{}
  \newcommand{\yifan}[1]{}
  \newcommand{\yunzhe}[1]{}
  \newcommand{\sitao}[1]{}
\else
  \newcommand{\todo}[1]{{\color{red}[TODO: #1]}}
  \newcommand{\ye}[1]{{\color{red}[Ye: #1]}}
  \newcommand{\zhiheng}[1]{{\color{red}[Zhiheng: #1]}}
  \newcommand{\yian}[1]{{\color{red}[Yian: #1]}}
  \newcommand{\yifan}[1]{{\color{red}[Yifan: #1]}}
  \newcommand{\yunzhe}[1]{{\color{red}[Yunzhe: #1]}}
  \newcommand{\sitao}[1]{{\color{red}[Sitao: #1]}}
\fi

\begin{document}

\title{COBRA: Algorithm-Architecture \underline{Co}-optimized \underline{B}inary T\underline{r}ansformer \underline{A}ccelerator for Edge Inference
}


\author{
    \IEEEauthorblockN{Ye Qiao\textsuperscript{\S}, Zhiheng Chen\textsuperscript{\S}, Yian Wang, Yifan Zhang, Yunzhe Deng, Sitao Huang} \\
    \textit{Department of Electrical Engineering and Computer Science} \\
    \textit{University of California,} Irvine, USA \\
    \{yeq6, zhihenc5, yianw11, yifanz58, yunzhd1, sitaoh\}@uci.edu
    \thanks{\textsuperscript{\S}Equal contribution}
}
\maketitle

\begin{abstract}

Transformer-based models have demonstrated superior performance in various fields, including natural language processing and computer vision. However, their enormous model size and high demands in computation, memory, and communication limit their deployment to edge platforms for local, secure inference. Binary transformers offer a compact, low-complexity solution for edge deployment with reduced bandwidth needs and acceptable accuracy. However, existing binary transformers perform inefficiently on current hardware due to the lack of binary specific optimizations. To address this, we introduce COBRA, an algorithm-architecture co-optimized binary Transformer accelerator for edge computing. COBRA features a real 1-bit binary multiplication unit, enabling matrix operations with -1, 0, and +1 values, surpassing ternary methods. With further hardware-friendly optimizations in the attention block, COBRA achieves up to 3,894.7 GOPS throughput and 448.7 GOPS/Watt energy efficiency on edge FPGAs, delivering a 311× energy efficiency improvement over GPUs and a 3.5× throughput improvement over the state-of-the-art binary accelerator, with only negligible inference accuracy degradation.

\end{abstract}

\maketitle

\section{Introduction}



In recent years, transformer-based models have become foundational architectures across multiple domains, achieving state-of-the-art performance in tasks such as natural language processing \cite{kenton2019bert}, computer vision \cite{alexey2020image}, and others \cite{kaeley2023support, singh2023progprompt}. These models excel at capturing complex patterns through self-attention mechanisms and extensive parametrization, often involving billions of parameters to achieve superior performance. However, the increasing size of these models introduces significant computational, memory, and communication challenges, restricting their deployment to a wide range of devices, especially resource-constrained devices. For many edge applications, such resource demands are impractical, making the deployment of full-scale transformers on edge devices particularly difficult, especially in scenarios requiring real-time inference.

To address these challenges, model compression techniques like quantization \cite{zafrir2019q8bert} and pruning \cite{kwon2022fast} have been proposed. Quantization reduces computational requirements by lowering numerical bitwidths, though it can lead to performance degradation. Despite these advances, quantized models often remain inefficient on edge devices due to the lack of effective acceleration on existing hardware. 

Binary transformers, an extreme form of quantized transformers, represent all weights and activations using binary values like $\{-1, 1\}$ or $\{0, 1\}$. These models significantly reduce computational complexity, model size, and communication bandwidth by replacing expensive floating-point operations with efficient bitwise logical operations \cite{zhirubnn, fracbnn, qiao2022two, 10025006}. This approach enables faster execution with lower power consumption, making binary transformers a promising solution for energy efficient edge deployment.

Despite these advantages, binary transformer models face significant challenges in implementation on existing hardware. The primary issue is the lack of optimized hardware units for binary matrix multiplication, as most processors and accelerators are designed for integer and floating-point computation, which are inefficient for binary operations. Furthermore, previously proposed binary transformer models are often not well-suited for hardware optimization. For instance, BinaryBERT \cite{bai2020binarybert} and BiBERT \cite{qin2022bibert} either do not fully binarize the model or retain some ternary representations. Similarly, BiT \cite{liu2022bit} employs both $\{-1, 1\}$ and $\{0, 1\}$ binarization schemes (though not for the same operation) and uses the original softmax function, which imposes a significant performance burden on hardware acceleration design, as will be discussed in a later section.


To overcome these limitations, we propose \textbf{COBRA}, a hardware/software co-designed Binary Transformer Accelerator optimized for edge FPGAs. COBRA introduces several innovations, including the \textit{Shifted Polarized Softmax (SPS)} for hardware-efficient attention, a true 1-bit binary multiplication method for \(-1\), \(0\), and \(+1\) matrices, and integer packing to maximize bandwidth. Additional optimizations, such as efficient popcount units, operation fusion, processing element reuse, and parallelism tuning, further enhance throughput, latency, and resource efficiency. Our evaluations show that COBRA operates efficiently on both mid-range edge FPGA (ZCU102) and low-power edge FPGA (KV260), making it portable and well-suited for resource-limited edge deployment on various devices.
The {contributions} of this work are summarized as follows:
\begin{itemize}

    \item We propose \textbf{COBRA}, an algorithm-architecture co-optimized hardware accelerator for efficient inference of binary Transformer models. COBRA achieves $3.5\times$ speedup over state-of-the-art binary transformer accelerators. 

    \item At hardware architecture level, we propose a \emph{real 1-bit} binary matrix multiplication engine, \textbf{\textit{RBMM}}. 
    It achieves high computational efficiency with real 1-bit operations, using bitwise XNOR and popcount operations with a unique ``don't care" (DC) count mechanism. RBMM is designed to support serveral variations of binary operations, making it reusable and resource-efficient for edge devices, and it could be extended to other binary models. 

    \item At algorithm level, we propose a hardware-friendly binary attention mechanism, shifted polarized softmax, \textbf{\textit{SPS}}, which enables efficient hardware implementation with negligible impact on transformer models' inference accuracy.

    \item At system level, we propose a series of optimizations including 6:3 compressor-based popcount, quantization-fused multiplication, and long bitwidth datapack representation that further enhance the system performance of COBRA.

\end{itemize}

\section{Background and Related Work}
\subsection{Binary Transformer}

\subsubsection{Binary Transformer Basic}

The binarized Transformer represents an extreme case of transformer quantization, where full-precision real number weights and activations \( W_r, A_r \) are approximated with binary values \( W_b, A_b \) as follows:
\[
W_r, A_r \approx \alpha W_b, \gamma A_b
\]
where \( W_b = \text{Sign}(W_r) \), \( \alpha = \frac{1}{n} \| W_r \| \), \( n \) is the total number of elements in \( W_r \). Activations are approximated with a similar way. Here, \( \alpha \) and \( \gamma \) are scaling factors, which may vary across different approaches. For example, the state-of-the-art binary transformer BiT \cite{liu2022bit} uses these scaling factors in the initialization, and further trains them as part of the model weights during updates. Our work adopts a similar strategy. Without considering bias, the originial full-precision matrix multiplication can be approximated with binary operations as:
\[
W_r \otimes A_r \approx \text{Popcount}\left(\text{XNOR}(W_b, A_b)\right) \alpha \gamma
\]
where $\otimes$ is matrix multiplication operator and popcount returns the number of ``$1$''s in a binary format number.
\subsubsection{Binary Transformer Designs}
Most transformer quantization efforts focus on 8-bit or 4-bit integer quantization due to the inherent difficulty of quantizing the attention block. Only a few studies address extreme low-bit transformers. For instance, TernaryBERT \cite{zhang2020ternarybert} proposed ternarizing the weights of the BERT model and fine-tuning it with a small dataset, achieving reasonable results. BinaryBERT \cite{bai2020binarybert} was the first to quantize both weights and activations of the BERT base model to binary, although it maintained a 2-bit embedding layer. This approach begins by initializing the network with weights from a fully trained half-sized ternary model, then proceeds by splitting weights to construct the binary model, followed by fine-tuning with knowledge distillation.

Built on BinaryBERT, two later works BiBERT \cite{qin2022bibert} and BiT \cite{liu2022bit} further binarized all weights, activations, and embeddings. They compensate performance loss and maximize representational capability through direct matching, precision-progressive distillation, a two-set binarization scheme, and an elastic binary activation function with learnable parameters. These efforts yielded substantial model accuracy. However, the lack of hardware considerations, such as efficient softmax implementation and compatibility with two-set binarization schemes, limits the ability of these models to be accelerated efficiently with hardware, posing challenges for specialized accelerator design.

\begin{figure}[t]
    \centering
    \includegraphics[width=0.9\linewidth]{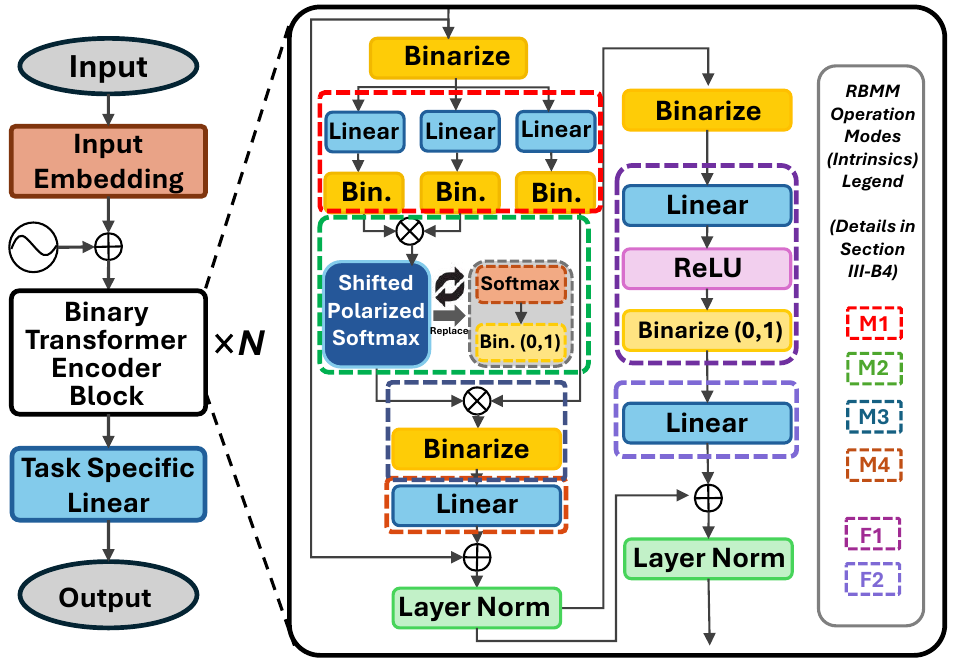}
    \caption{General structure of binarized BERT with our Shifted Polarized Softmax (SPS) and Real Binary RBMM Engine}
    \label{fig:Binary_Transformer}
\end{figure}

\subsection{Hardware Acceleration of Binary Transformers}

\subsubsection{Binary CNN Accelerators Are Not Suitable for Transformers} While previous works have proposed accelerators for binary convolutional neural networks (CNNs) \cite{fracbnn, andri2017yodann, al2018xnorbin}, they are unsuitable for binary transformers due to differences in computation irregularity, data dependency, and memory access patterns. Transformers rely on fine-grained parallelism across long sequence dimensions for self-attention and large matrix multiplications, unlike the structured parallelism of CNNs. Furthermore, binary transformers require two specific types of  binary schemes to function effectively and achieve acceptable accuracy---a challenge not present in binary CNNs.

\subsubsection{Binary Transformer Acceleration}
Hardware acceleration for binary transformer networks presents unique challenges compared to full-precision networks. Fig. \ref{fig:Binary_Transformer} illustrates the general structure of a binarized BERT model. Binary Transformer models require two types of binarization schemes, \(\{-1, 1\} \otimes \{-1, 1\}\) and \(\{-1, 1\} \otimes \{0, 1\}\), and involve two distinct matrix multiplication operations: linear layer and attention computation. 

Some existing works aim to enhance the efficiency of binary transformers. VAQF \cite{sun2022vaqf} introduced a binary vision Transformer (ViT) accelerator that exploits the speedup potential of binarization by converting multiplication into bitwise logical operations, as discussed in an earlier section. However, it only supports one execution mode and does not consider the matrix multiplication of binarized attention. BETA \cite{ji2024beta} employs a Compressor Tree Loop to create an accelerator specifically for fully binary transformers, supporting both matrix multiplication types and achieving improved efficiency and throughput on edge platforms. BAT \cite{ji2024co} offers a co-design approach, creating a custom binarized transformer model with a specialized hardware accelerator for edge deployment. However, they rely on encoding ternary values \(\{-1, 0, 1\}\) as two-bit numbers through lookup tables, which is not a real binary representation, therefore it did not fully leverage the unique advantages of binary models. 

In summary, no existing hardware acceleration works fully leverage the potential of binary transformers, nor address the primary performance (latency) bottleneck for binary accelerator design: the softmax operations. Our proposed \textbf{\textit{RBMM}} engine, along with the \textbf{\textit{Shifted Polarized Softmax (SPS)}}, addresses these limitations and unlocks the full potential of binary transformers for edge deployment.

\subsubsection{Attention Mechanism Acceleration}
The self-attention, constitutes the cornerstone of transformer models, enabling the dynamic weighting of input sequence elements based on their relative importance. However, this capability incurs a significant computational cost: attention mechanisms exhibit quadratic time and space complexity relative to sequence length. Flash Attention \cite{flash_attention} represents a prominent GPU-optimized approach to accelerating attention computations. Its architecture leverages HBM and SRAM on GPUs, employing a tiled computation strategy that integrates the softmax and matrix multiplication operations. Nevertheless, this method is not directly adaptable to edge FPGAs, which are constrained by the limited bandwidth of DDR memory and limited MB-level on-chip BRAM and URAM.
\section{COBRA Methodology}

\begin{figure}[t]
    \centering
    \includegraphics[width=0.9\linewidth]{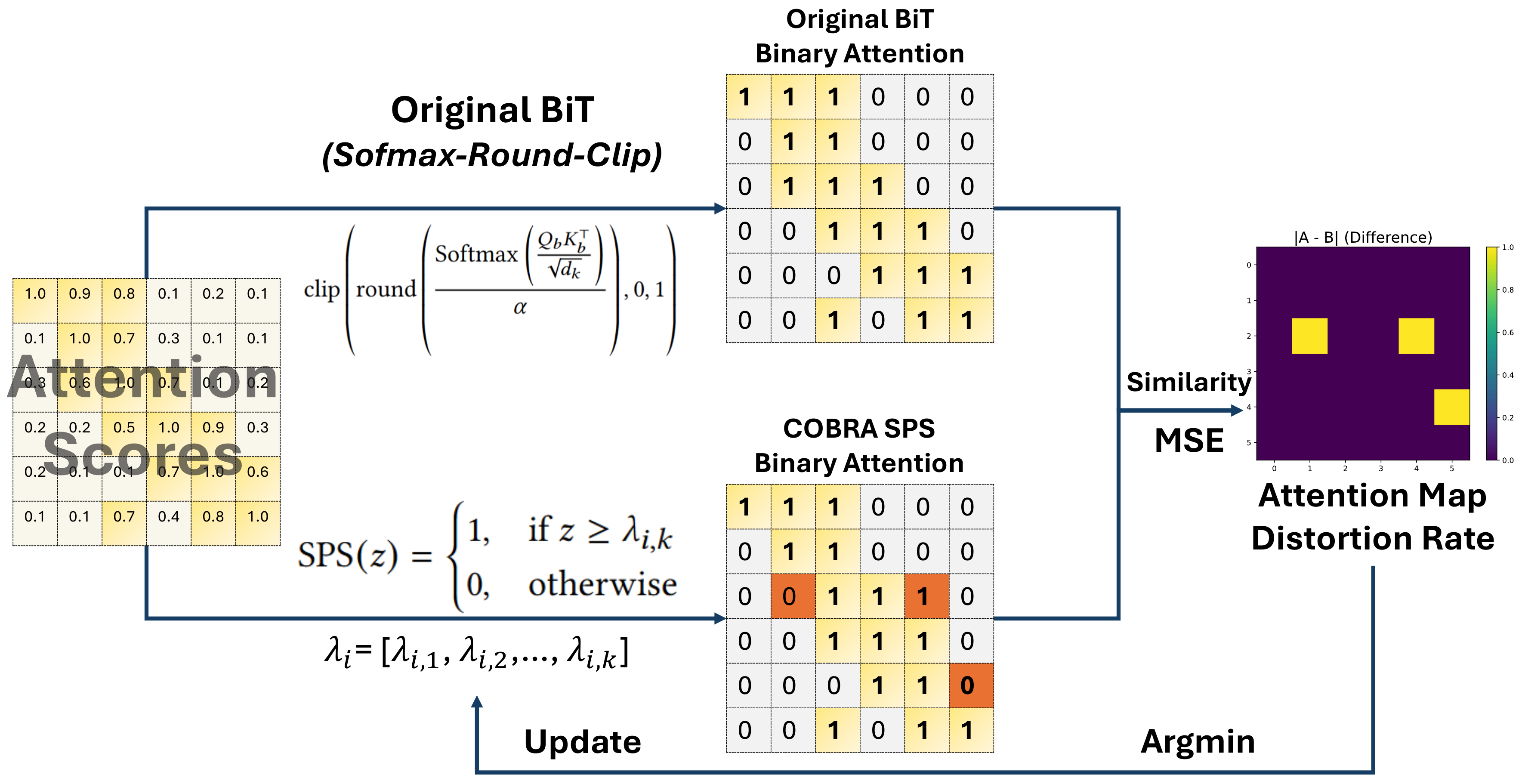}
    \caption{Shifted Polarized Softmax (SPS) Search}
    \label{fig:SPS}
\end{figure}

\subsection{Hardware Friendly Binary Attention Mechanism Design}
\renewcommand{\arraystretch}{1}
\begin{table*}[!ht]
    \centering
    \caption{Accuracy of COBRA vs. Others. All binary transformer models use W1A1 (1-bit weights and activations).}
    \label{tab:accuracy-result}
    \resizebox{0.9\textwidth}{!}{
    \begin{tabular}{l c c c c c c c c c c}
        \toprule
        \textbf{Variations}   & $\textbf{MNLI}_{\text{\scriptsize -m/mm}}$ & \textbf{QQP} & \textbf{QNLI} & \textbf{SST-2} & \textbf{CoLA} & \textbf{STS-B} & \textbf{MRPC} & \textbf{RTE} & \textbf{Average} & \textbf{Relative Perf.} \\
        \midrule
        Original BERT\cite{kenton2019bert}    & 84.9/85.5 & 91.4 & 92.1 & 93.2 & 59.7 & 90.1 & 86.3 & 72.2 & 83.9 & 129.1\%\\
        \midrule
        BinaryBERT\cite{bai2020binarybert} & 35.6/35.3 & 66.5 & 51.5 & 53.2 & 0 & 6.1 & 68.3 & 52.7 & 41.0 & 55.6\%\\
        BiBERT\cite{qin2022bibert}  & 69.5/71.1 & 84.9 & 78.3 & 87.5 & 25.4 & 34.7 & 72.5 & 49.8 &  63.8 & 89.7\%\\
        BiT\cite{liu2022bit}     & 75.6/76.3 & 84.7 & 82.8 & 88.0 & 27.4 & 69.8 & 77.4 & 52.7 & 71.0 & 100\% \\
        \midrule 
        {COBRA (Layer)}  & {67.6/68.7} & {79.03} & {70.1} & {85.7} & {25.4} & {66.3} & {74.3} & {53.4} & {65.1} & 93.8\% \\ 

        {COBRA (Row)}  & {-} & {-} & {71.5} & {87.1} & {26.6} & {-} & {-} & {53.0} & {-} & {-} \\ 

        \textbf{COBRA (Head)}  & \textbf{71.3/72.1} & \textbf{83.3} & \textbf{73.2} & \textbf{86.9} & \textbf{27.9} & \textbf{69.2} & \textbf{80.2} & \textbf{53.4} & \textbf{68.2} & \textbf{98.2\%}\\ 

        \bottomrule
    \end{tabular}
    }
    \vspace{2mm}
    \caption*{\footnotesize{* BiT~\cite{liu2022bit} was selected as the baseline, as it represents the state-of-the-art in binary BERT models.}}
\end{table*}


\subsubsection{Real Binary Attention}
BiBERT \cite{qin2022bibert} and BiT \cite{liu2022bit} followed Eq. (\ref{alg:bsoftmax}) to implement softmax in their binary attentions. This approach overlooks hardware-level overhead of softmax, leading to suboptimal computation speed, hardware costs, and power efficiency, making it impractical for edge deployment. Furthermore, the use of softmax requires exponential computation and produces non-binary values in the attention mechanism, preventing BiT from achieving a fully binarized transformer model and complicating efficient hardware acceleration. In detail, BiT calculates a vanilla attention score (including softmax operation) and applies elastic binarization as Eq. (\ref{alg:bit}), effectively redistributing the attention score to binary domain. 
\begin{equation} \label{alg:bsoftmax}
   \text{Softmax}(z_i) = \frac{\exp(z_i)}{\sum_{j=1}^{n} \exp(z_j)}
\end{equation}

\begin{equation} \label{alg:bit}
    Att\_prob_{BiT} = \text{clip}\left(\text{round}\left(\frac{\text{Softmax}\left(\frac{Q_b K_b^\top}{\sqrt{d_k}}\right)}{\alpha}\right), 0, 1\right)
\end{equation}
where $\text{clip}(\cdot, 0, 1)$ clamps input elements into  $[0,1]$, and $\text{round}(\cdot)$ rounds input to the nearest integer.
However, both the softmax operation and the elastic binarization function rely on many floating-point operations, which dominate the computational complexity of binary models and significantly compromise the hardware acceleration.

To overcome these limitations, we propose Shifted Polarized Softmax (SPS) as a direct replacement for softmax and binarization in attention computation. SPS generates binary-state outputs directly, without relying on any floating-point arithmetic, thereby eliminating the need for a separate binarization step and further improving hardware efficiency.


\subsubsection{Shifted Polarized Softmax (SPS)} \label{sec:sps}


Conventional softmax scales input values into a continuous probability distribution over the range (0,1) and the quantization function rescale the distribution back to binary form, the SPS function instead polarizes values directly into binary states based on a learnable threshold, capturing the core purpose of softmax—emphasizing stronger signals—without the need for continuous probabilistic scaling, as shown in Eq. (\ref{alg:sps}). 
\begin{equation} \label{alg:sps}
   \text{SPS}(z) = 
   \begin{cases}
       1, & \text{if } z \geq \lambda_{i,k} \\
       0, & \text{otherwise}
   \end{cases} 
\end{equation}
where $i$ indicate $i^{th}$ attention block and $k$ represent $K^{th}$ attention head in the attention block.
\begin{equation} \label{alg:COBRA}
    Att\_prob_{SPS} = \text{SPS}\left(\frac{Q_b K_b^\top}{\sqrt{d_k}}\right)
\end{equation}
As discussed in \cite{qin2022bibert}, for the attention mechanism to capture crucial elements in the binary representation, maximizing the information entropy of binarized attention score is a must and a binarized attention score matrix can serve the purpose. Hence, we propose SPS to replace Eq. (\ref{alg:bit}) by search for a series of head-wise thresholds $\lambda \in h\times n$ that further compensate the information loss for not having softmax normalization for attention score. SPS approximates the effects of softmax plus elastic binarization function while significantly reducing computational overhead, making it more compatible with hardware constraints.
Our Final attention probability can be found as Eq. (\ref{alg:COBRA}). 

\subsubsection{SPS Thresholds Search and Model Training} \label{sec:sps_search}

\begin{figure}[t]
\centering

\begin{subfigure}[t]{0.5\textwidth}
    \centering
    \includegraphics[width=\textwidth]{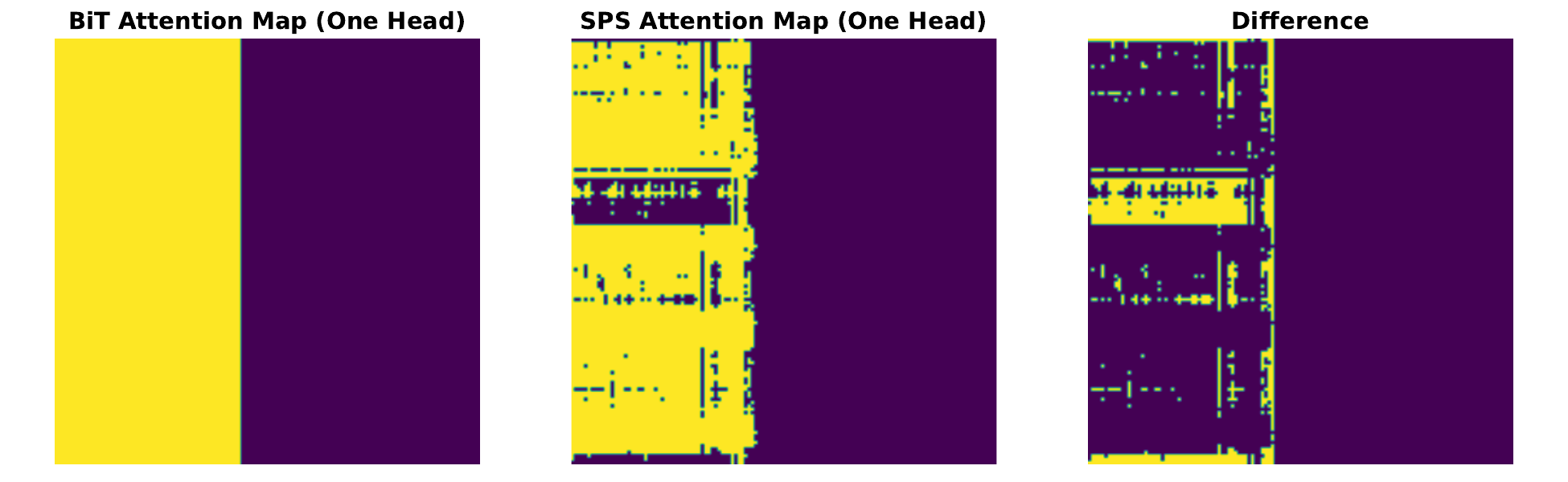}
    \caption{Heatmap Correlation Comparison}
\end{subfigure}
\vspace{4mm} 

\begin{subfigure}[t]{0.23\textwidth}
    \centering
    \includegraphics[width=\textwidth]{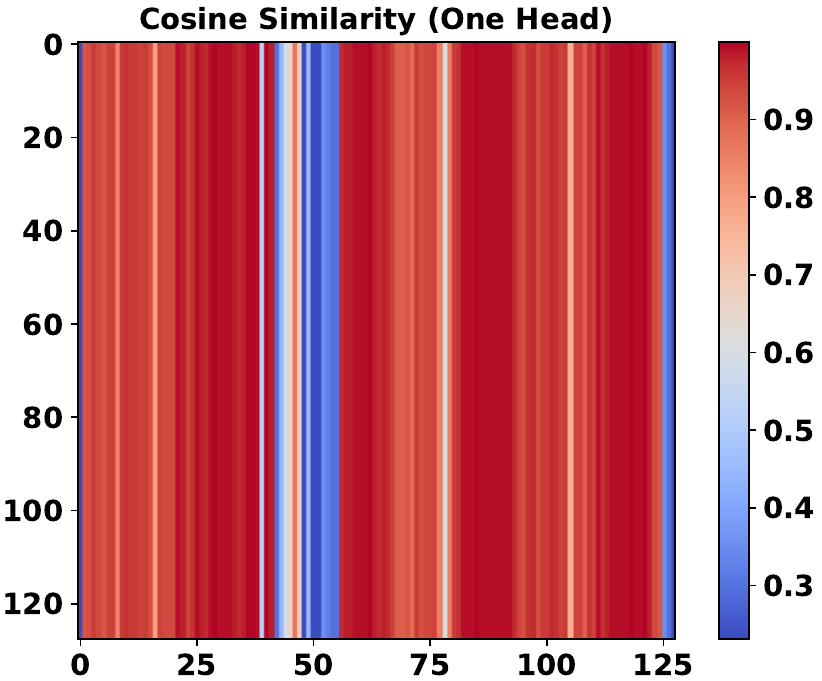}
    \caption{Cosine Similarity Comparison}
\end{subfigure}
\begin{subfigure}[t]{0.25\textwidth}
    \centering
    \includegraphics[width=\textwidth]{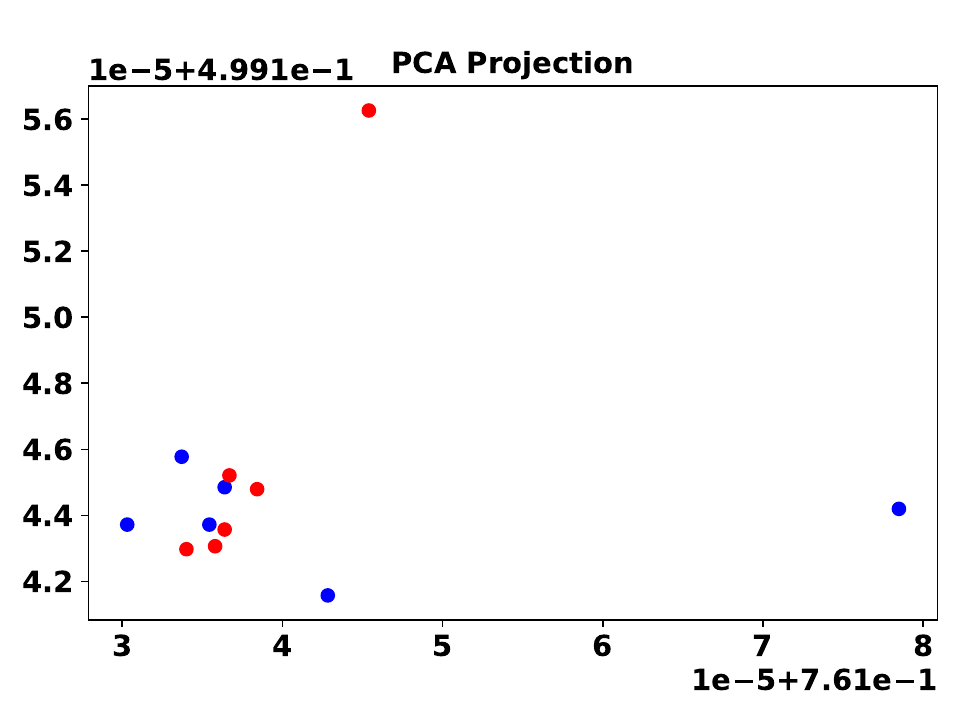}
    \caption{PCA Projection Comparison}
\end{subfigure}
\vspace{4mm}

\begin{subfigure}[t]{0.49\textwidth}
    \centering
    \includegraphics[width=\textwidth]{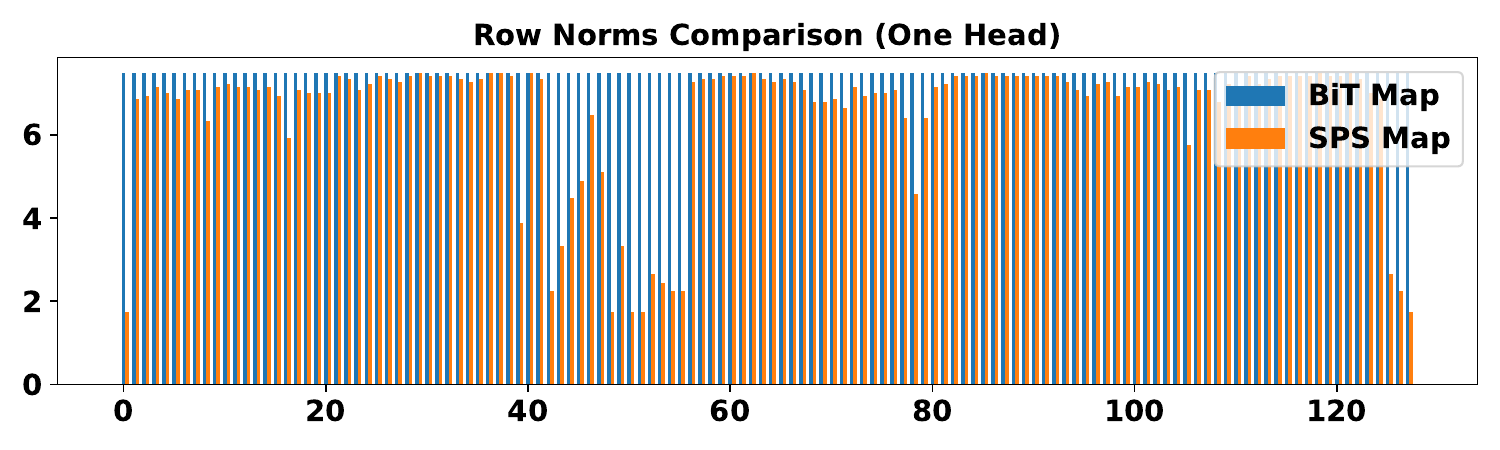}
    \caption{Attention Row Norm Comparison}
\end{subfigure}

\caption{Similarity and Correlation Comparisons Between BiT (with Regular Softmax) and SPS Attention.}
\label{fig:SPS_Corralation}
\end{figure}

Figure~\ref{fig:SPS} illustrates our method and search strategy for determining the SPS thresholds. Suppose two attention maps, $A_1$ and $A_2$, are obtained from the same layer $Layer_i$. We define the \textbf{Channel Distortion Rate (CDR)} as the mean squared error (MSE) between these two maps:

\begin{equation}
    \text{Distortion} = \frac{1}{n} \sum_{n=1}^{i} (A_1 - A_2)^2
\end{equation}

The thresholds $\lambda \in \mathbb{R}^{h \times n}$ are predefined within the range $[0, 1]$ with a search granularity of 0.05 and an initial value of 0(regular sign function). Our objective is to identify the optimal thresholds that minimize the output discrepancy between BiT’s softmax-based attention (Eq.~\ref{alg:bit}) and our SPS-based attention (Eq.~\ref{alg:COBRA}) for a given attention layer. Formally, we solve:
\begin{small}
    
\vspace{-2mm}
\begin{equation}
\lambda^* = \arg\min_{\lambda} \left( \text{Distortion}\left(\text{Att\_prob}_{\text{BiT}},\ \text{Att\_prob}_{\text{SPS}}(\lambda) \right) \right)
\vspace{-1mm}
\end{equation}
\end{small}
Here, $\lambda$ represents the per-head threshold. To avoid overfitting to any specific task, we uniformly sample 10\% of each dataset (benchmark) to construct a small calibration set for the search process. This process identifies a unique threshold for each attention head in every attention block and is highly efficient, requiring only about five minutes in our experiments.

We further evaluate different threshold granularities—per layer, per head, and per row of the attention map—which will be discussed in the next section. Balancing effectiveness and efficiency, we adopt the head-wise threshold configuration in our final design. This decision is well-supported, as different attention heads often capture distinct relational patterns among tokens and contribute to diverse representations, thereby enhancing the model’s expressivity.

Figure~\ref{fig:SPS_Corralation} presents multiple similarity and correlation measurements between BiT (with Softmax) and SPS attention, demonstrating the validity of our SPS approximation. Finally, to further enhance COBRA’s performance, we fix the searched thresholds $\lambda$ and fine-tune the model weights using the original training data to compensate for the information loss introduced by the SPS approximation.

\subsubsection{SPS Results Analysis} \label{sec:sps_results}

Table \ref{tab:accuracy-result} presents the accuracy performance of our binary COBRA model on the GLUE benchmarks \cite{glue}, compared against the original BERT and other binary BERT variants. Our model outperforms both BinaryBERT \cite{bai2020binarybert} and BiBERT \cite{qin2022bibert} by a significant margin, with only a less than 2\% minor average performance drop compared to the state-of-the-art BiT \cite{liu2022bit}. Despite this small accuracy reduction, our model relies exclusively on true binary operations and provides substantial hardware advantages, which will be further discussed in Section \ref{subsec:ablation}.

We also evaluate different granularities of the SPS threshold. The attention layer-wise threshold performs adequately, with minimal overhead and the shortest search time. However, as previously discussed, the head-wise threshold yields the best performance, incurring similarly negligible overhead compared to the layer-wise case, and the search can be completed relatively quickly. Additionally, we conducted experiments using a more fine-grained row-wise threshold at the attention map. The results indicate that it does not offer meaningful performance improvements while introducing more parameters to the model. It also increases the search time by over 20×. Therefore, we adopt the head-wise threshold as the default configuration for the SPS function in our COBRA models. Additionally, all threshold granularity configurations are fully supported by our proposed accelerator architecture without incurring any extra computation time.

\subsection{COBRA Hardware Accelerator Design}

\subsubsection{Real 1-bit Binary Matrix Multiplication (RBMM)}
\begin{figure}[t]
    \centering
    \includegraphics[width=0.9\linewidth]{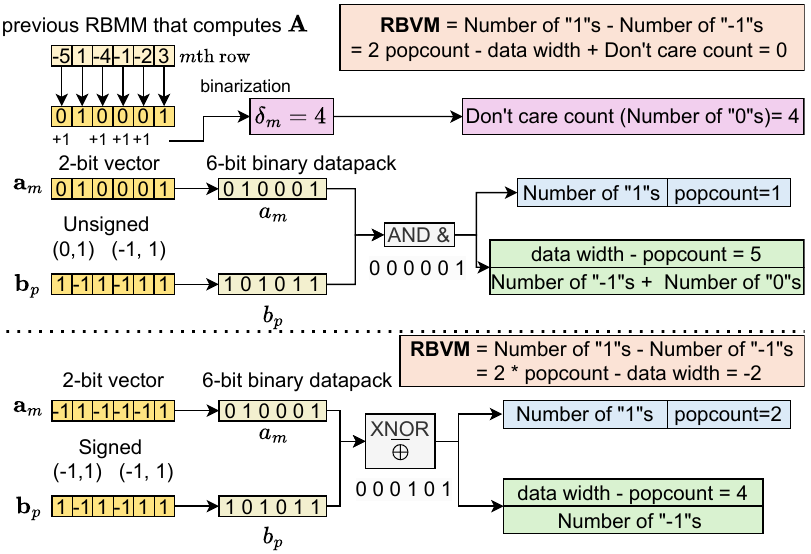}
    \caption{A 6-bit example of our RBVM.}
    \label{fig:RBVM}
\end{figure}
To fully leverage the benefits of binary representations, we propose an efficient algorithm, \textbf{RBMM}, for real binary matrix multiplication.  
Consider matrices ${\bf A} = \left[{\bf a}_1, \cdots, {\bf a}_{m}, \cdots, {\bf a}_{M} \right]^{\top} \in {\{(0, 1) | (-1, 1)\}^{M \times N}}$, ${\bf B} = \left[{\bf b}_1, \cdots,{\bf b}_{p}, \cdots, {\bf b}_{P} \right] \in {\{(-1, 1)\}^{N \times P}}$, and multiplication result ${\bf C} \! = \!{\bf A} \otimes {\bf B} \in \mathbb{N}^{M \times P}$. 
\begin{figure}[t]
    \centering
    \includegraphics[width=0.9\linewidth]{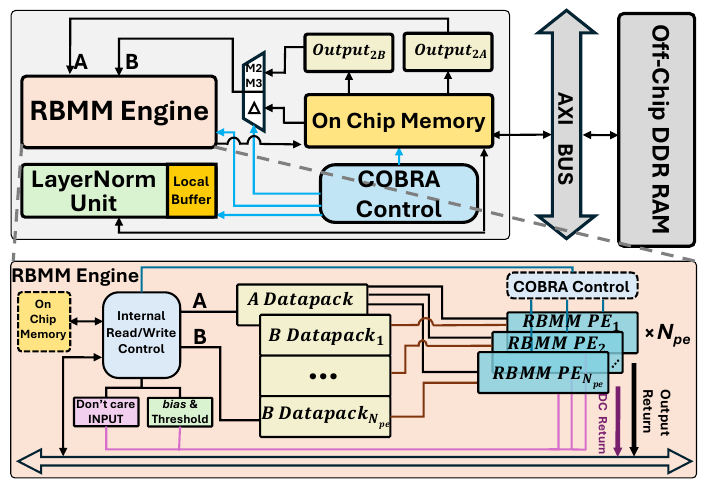}
    \caption{COBRA Hardware Architecture Overview.}
    \label{fig:Architecture_all}
\end{figure}
To optimize binary computations while eliminating sign bit management in ternary multiplication, we propose a unified representation for matrices $\bf A$ and $\bf B$, where ``$-1$" is encoded as ``$0$" while preserving ``$1$". For computational efficiency, the bit-packing strategy is adopted where elements ${\bf a}_{m}$ are packed into $N$-bit $a_m$ datapack, and similarly, ${\bf b}_{p},$ are packed into $N$-bit $b_{p}$.
Then, the real binary vector dot product (RBVM), denoted as ${\bf a}_m{\bf b}_p = a_m \boxtimes b_m$, can be derived as (steps omitted for brevity):
\begin{align} \label{alg:xnor}
a_m \!\boxtimes\! b_p\! = \!
\begin{cases} 
2 \, \text{popcount}(a_m \overline{\oplus} b_p)\! - \!N, & a_m \! \in \!\{(-1, 1)\}\\
2 \,\text{popcount}(a_m \& b_p) \! - \! N \!+ \!{\delta}_{m}, & a_m \!\in \! \{(0, 1)\}\\
\end{cases}
\end{align}
where $N$ represents the data bitwidth of the datapack, popcount counts the number of ``1''s in the vector,  $\overline{\oplus}$  and $\&$ denote bit-wise logical operations XNOR and AND respectively, and ${\delta}_{m}$ denotes the number of zeros in the unsigned datapack $a_m$ (``don't care'' count, DC). Fig. \ref{fig:RBVM} illustrates our RBVM algorithm with a 6-bit toy example (768 bits in our actual design). The algorithm determines the count of ``$1$"s through popcount operations, from which the number of ``$-1$"s can be derived. For the $(0, 1)$ binarization scheme, the number of ``0"s are considered. Moreover, $-N$ can be fused with ``$\geq T$'' binarization in Eq. (\ref{alg:sps}), which will be detailed in the Section \ref{sec:RBMM}.
Although this approach seemingly requires an additional popcount operation to determine $\delta_{m}$, this computation can be efficiently integrated into the previous quantization-fused RBMM as shown in Fig. \ref{fig:RBVM}. Specifically, when a ``0" quantization value is calculated, the row's $\delta_{m}$ is incremented by 1. Besides, the RBVM operation exhibits compositional properties when column/row vectors are subdivided into $S$ smaller vectors, where ${\bf a}_{m} = [{\bf a}_{m_1}, ...,{\bf a}_{m_s}, ... ,{\bf a}_{m_S}]$ and ${\bf b}_{p} = [{\bf b}_{p_1}, ...,{\bf b}_{p_s}, ... ,{\bf b}_{p_S}]^{\top}$. The vector dot product property is preserved as follows,
\begin{align}
{\bf a}_{m} {\bf b}_{p} = \sum_{s=1}^{S} {\bf a}_{m_s}{\bf b}_{p_s} = \sum_{s=1}^{S} { a}_{m_s} \boxtimes { b}_{p_s}, \label{ACC_HEAD_PE}
\end{align}
where $a_m = \text{concat}(a_{m_1}, ...,a_{m_s}, ..., a_{m_S})$ and $b_p = \text{concat} (b_{p_1}, ...,b_{p_s}, ..., b_{p_S})$. This decomposition enables the RBMM engine to support both smaller-sized RBMM operations and larger configurations, making the RBMM engine reusable for both multi-head attention (MHA) and feed-forward network (FFN). 

\subsubsection{Quantization-fused RBMM} On top of our RBMM design, we further fuse quantization into it and create the quantization-fused RBMM engine. The unsigned and signed binary quantization schemes can then be mathematically expressed as:
\begin{align}
c^{b}_{ij} = 
\begin{cases} 
\text{clip}\left(\text{round}\left(\frac{c_{ij}-\beta_{j}}{\alpha}\right), 0, 1\right), &\text{(0, 1) scheme}\\
\text{sign}\left(\frac{c_{ij}-\beta_{j}}{\alpha}\right),  &\text{(-1, 1) scheme}\\
\end{cases}
\end{align}
where $c_{ij}$ is an element from $\bf C$, $c_{ij}^{b}$ is the binarized value, $\alpha \in \mathbb{N}^{+}$ denotes the scaling factor, $\beta_{i}$ is an element from the shift vector $\boldsymbol{\beta} \in \mathbb{N}^{1 \times P}$ and the $sign$ of zero is deemed as ``$1$''. Since the unified hardware representation for ``$-1$" and ``$0$" has been applied as mentioned earlier, the two binarization schemes can be unified as,
\begin{align} \label{alg:bias}
c^{b}_{ij} = 
\begin{cases} 
1, c_{ij}-\theta_{j}\geq 0 \\
0, c_{ij}-\theta_{j}<0\\
\end{cases}
\!\!\!\!\!,  \,\,\theta_{j}=
\begin{cases} 
r\left(\frac{1}{2}\alpha + \beta_{j}\right), \!&\text{(0, 1)}\\
\beta_{j}, &\text{(-1, 1)}\\ 
\end{cases}
\end{align}
where $\theta_{j}$ is an element from the $bias$ vector, and $r$ is the round operation.
Furthermore, the unsigned binarization is only employed after the ReLU operation within the feed-forward network (FFN), as illustrated in Fig. \ref{fig:Architecture_all}. From a hardware implementation perspective, these operations can be efficiently merged due to the overlapped range between the conditions $c_{ij}\geq r(\frac{1}{2}\alpha + \beta_{j})$ and $c_{ij}\geq 0$ in the ReLU function. Consequently, the $\theta_{j}$ is set as $\max (0, r(\frac{1}{2}\alpha + \beta_{j}))$.

\begin{figure}[t]
    \centering
    \includegraphics[width=\linewidth]{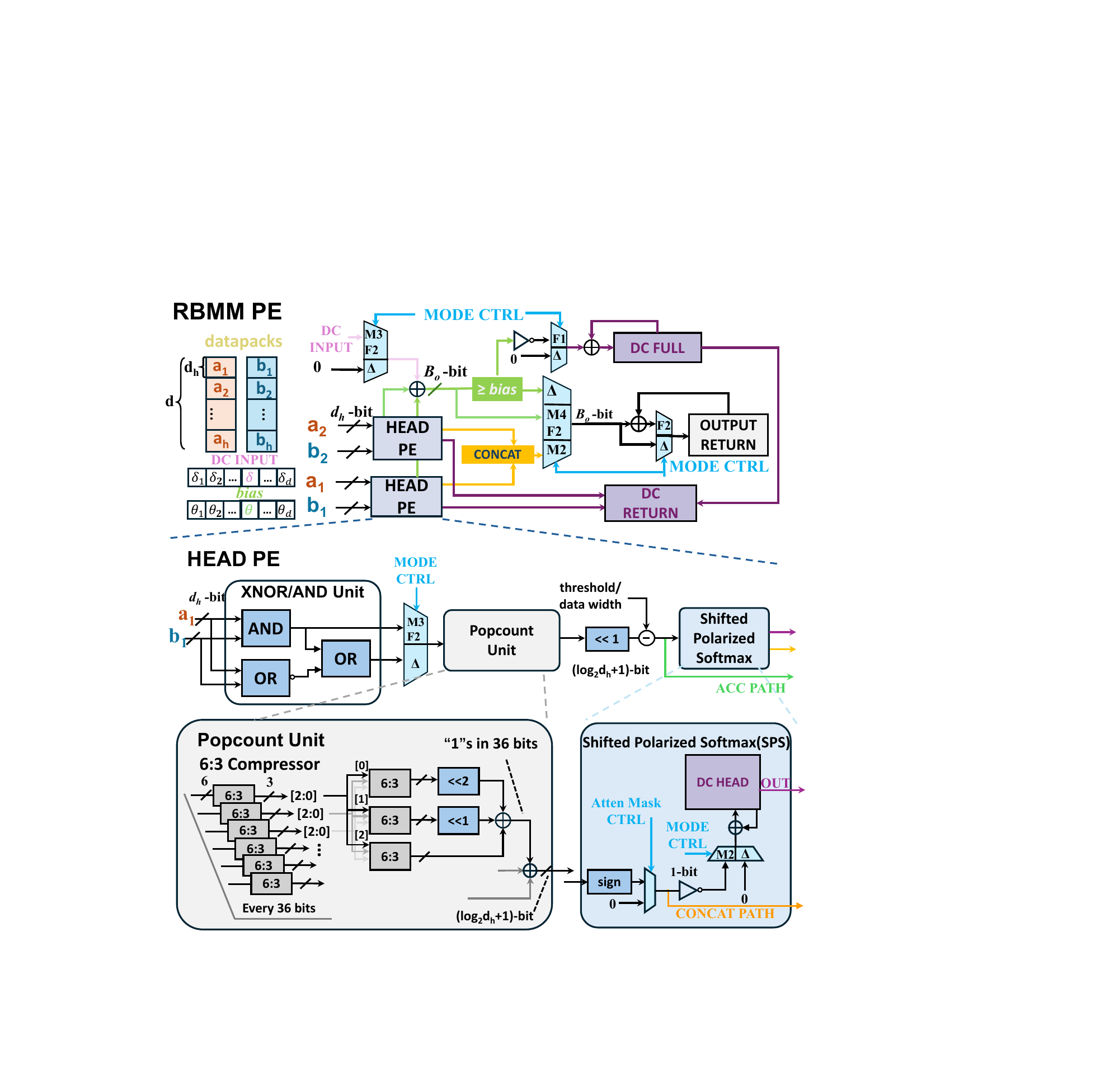}
    \caption{The Architecture of RBMM Processing Elements (M1--4  and F1--2 are operation modes of RBMM engine). }
    \label{fig:RBMM PE}
\end{figure}


\subsubsection{Overall Hardware Architecture}
As illustrated in Fig. \ref{fig:Architecture_all}, the accelerator architecture consists of five major components: an RBMM engine, a LayerNorm unit, a COBRA configuration control unit, a data packing conversion unit, and a master AXI interface.  For clarity, we define the following notation: $d$ represents the hidden dimension, $h$ denotes the number of heads, $d_h = d/h$ indicates the size per head, $FF_{size} = Rd$ ($R > 1$) represents the FFN hidden dimension, $l$ ($l < d$) denotes the sequence length, and $N_{pe}$ represents the number of Processing Elements (PEs). The numbers in the multiplexer indicate the RBMM modes (detailed in the next Section \ref{sec:RBMM}), while $\Delta$ represents all other modes.  The notation ($n$ $x\times y \times z$) represents the RBMM where $n$ is the number of matrix multiplications, the first matrix has dimensions $x \times y$, the second matrix has dimensions $y \times z$, and the result matrix has dimensions $x \times z$.

The RBMM engine supports multiple configurations (i.e., operation modes) of quantization-fused matrix multiplication operations. We will discuss the operation modes of RBMM in the next Section \ref{sec:RBMM}. The LayerNorm unit implements the normalization function using the 16-bit fixed-point scaling factors and value for computation to optimize hardware efficiency. The COBRA control unit manages the RBMM configuration parameters and orchestrates the overall computation flow. The data packing conversion unit efficiently transforms intermediate matrices into binary datapack formats. External DDR memory accesses like weight matrix fetching and intermediate array storage are facilitated via the AXI high-performance master interface. 

\begin{table*}[!ht]
  \centering
  \caption{COBRA Hardware Performance Evaluation and Comparison against Previous Works}
\label{tab:performance}
\resizebox{\textwidth}{!}{
\begin{tabular}{c|ccc|cccccccccll}
\toprule
\multirow{3}{*}{Platform}                                             & \multicolumn{3}{c|}{GPU}                  & \multicolumn{11}{c}{FPGA}                                                                                                                                                                                                                                                                                                                                                     \\ \cline{2-15} 
                                                                      & \multicolumn{3}{c|}{RTX 3090}             & \multicolumn{1}{c|}{FQ-BERT\cite{FQ-BERT}} & \multicolumn{1}{c|}{COSA\cite{COSA}}    & \multicolumn{1}{c|}{TransFRU\cite{TransFRU}}   & \multicolumn{3}{c|}{BETA\cite{ji2024beta}}                           & \multicolumn{1}{c|}{BAT\cite{ji2024co}}     & \multicolumn{1}{c|}{\textbf{KV260}}                                                         & \multicolumn{3}{c}{\textbf{ZCU102}}                                                        \\ \hline
Network                                                               & BERT                & BiT                 & COBRA              & \multicolumn{1}{c|}{FQ-BERT} & \multicolumn{1}{c|}{BERT}    & \multicolumn{1}{c|}{BERT}       & BiT     & BinaryBERT & \multicolumn{1}{c|}{BiBERT}  & \multicolumn{1}{c|}{BMT}     & \multicolumn{1}{c|}{\textbf{\begin{tabular}[c]{@{}c@{}}COBRA \\ $N_{pe}=16$\end{tabular}}} & \multicolumn{3}{c}{\textbf{\begin{tabular}[c]{@{}c@{}}COBRA \\ $N_{pe}=32$\end{tabular}}} \\ \hline
Precision                                                             & W16A16                & W1A1                & W1A1                & \multicolumn{1}{c|}{W4/8A8}  & \multicolumn{1}{c|}{W8A8}    & \multicolumn{1}{c|}{W4A4}       & W1A1    & W1A1       & \multicolumn{1}{c|}{W1A1}    & \multicolumn{1}{c|}{W1A2}    & \multicolumn{1}{c|}{\textbf{W1A1}}                                                          & \multicolumn{3}{c}{\textbf{W1A1}}                                                          \\ \hline
\begin{tabular}[c]{@{}c@{}}Throughput\\ (GOPS)\end{tabular}           & 749.1               & 484.26              & 503.14              & \multicolumn{1}{c|}{22.74}   & \multicolumn{1}{c|}{1556.5}  & \multicolumn{1}{c|}{1223.90}    & 1240.98 & 1387.59    & \multicolumn{1}{c|}{1436.07} & \multicolumn{1}{c|}{1122.40} & \multicolumn{1}{c|}{\textbf{846.28}}                                                        & \multicolumn{3}{c}{\textbf{3894.74}}                                                       \\ \hline
\begin{tabular}[c]{@{}c@{}}Power \\ (W)\end{tabular}                  & 350.00              & 350.00              & 350.00              & \multicolumn{1}{c|}{9.80}    & \multicolumn{1}{c|}{31.3}    & \multicolumn{1}{c|}{45.46}      & 7.18    & 7.95       & \multicolumn{1}{c|}{8.20}    & \multicolumn{1}{c|}{5.47}    & \multicolumn{1}{c|}{\textbf{4.07}}                                                          & \multicolumn{3}{c}{\textbf{8.68}}                                                          \\ \hline
\begin{tabular}[c]{@{}c@{}}Energy Efficiency \\ (GOPS/W)\end{tabular} & 2.14                & 1.38                & 1.44                & \multicolumn{1}{c|}{2.32}    & \multicolumn{1}{c|}{49.7}    & \multicolumn{1}{c|}{26.9}       & 172.41  & 174.59     & \multicolumn{1}{c|}{175.23}  & \multicolumn{1}{c|}{207.7}   & \multicolumn{1}{c|}{\textbf{207.93}}                                                        & \multicolumn{3}{c}{\textbf{448.70}}                                                        \\ \bottomrule
\end{tabular}}
\end{table*}

\begin{table}[!ht]
 \centering
\caption{FPGA Resource Utilization Comparison. ``--" means not reported or don't have the kind of resource.}
    \label{tab:resource utilization}
 
    \resizebox{\linewidth}{!}{
\begin{tabular}{c|c|c|c|c|c|cc}
\toprule
Work     & FQ-BERT \cite{FQ-BERT}                                              & COSA \cite{COSA}                                                                & TransFRU \cite{TransFRU}                                               & BETA \cite{ji2024beta}                                                             & BAT \cite{ji2024co}                                                   & \multicolumn{2}{c}{\textbf{Our Work}}                                                                                                                               \\ \hline
Platform & ZCU102                                                & XCVU13P                                                               & Alveo U280                                             & ZCU102                                                            & ZCU102                                                 & \multicolumn{1}{c|}{\textbf{KV260}}                                                         & \textbf{ZCU102}                                                        \\ \hline
Network  & FQ-BERT                                               & BERT                                                                  & BERT                                                   & \begin{tabular}[c]{@{}c@{}}BiT\\ BiBERT\\ BinaryBERT\end{tabular} & BMT                                                    & \multicolumn{1}{c|}{\textbf{\begin{tabular}[c]{@{}c@{}}COBRA \\ $N_{pe}=16$\end{tabular}}} & \textbf{\begin{tabular}[c]{@{}c@{}}COBRA \\ $N_{pe}=32$\end{tabular}} \\ \hline
LUT      & \begin{tabular}[c]{@{}c@{}}123K\\ (45\%)\end{tabular} & \begin{tabular}[c]{@{}c@{}}1147K\\ (66\%)\end{tabular}               & \begin{tabular}[c]{@{}c@{}}633K\\ (49\%)\end{tabular} & \begin{tabular}[c]{@{}c@{}}191K \\ (70\%)\end{tabular}            & \begin{tabular}[c]{@{}c@{}}146K\\ (53\%)\end{tabular}  & \multicolumn{1}{c|}{\textbf{\begin{tabular}[c]{@{}c@{}}86K \\ (74\%)\end{tabular}}}         & \textbf{\begin{tabular}[c]{@{}c@{}}122K\\ (45\%)\end{tabular}}         \\ \hline
FF       & \begin{tabular}[c]{@{}c@{}}124K\\ (23\%)\end{tabular} & \begin{tabular}[c]{@{}c@{}}526K\\ (15\%)\end{tabular}                & \begin{tabular}[c]{@{}c@{}}---\\ (---)\end{tabular}    & \begin{tabular}[c]{@{}c@{}}88K\\ (16\%)\end{tabular}              & \begin{tabular}[c]{@{}c@{}}136K \\ (25\%)\end{tabular} & \multicolumn{1}{c|}{\textbf{\begin{tabular}[c]{@{}c@{}}105K\\ (45\%)\end{tabular}}}         & \textbf{\begin{tabular}[c]{@{}c@{}}162K\\ (30\%)\end{tabular}}         \\ \hline
DSP      & \begin{tabular}[c]{@{}c@{}}1751\\ (70\%)\end{tabular} & \textit{\begin{tabular}[c]{@{}c@{}}8192(est.)\\ (67\%)\end{tabular}} & \begin{tabular}[c]{@{}c@{}}5016\\ (56\%)\end{tabular}  & \begin{tabular}[c]{@{}c@{}}64 \\ (2.5\%)\end{tabular}              & \begin{tabular}[c]{@{}c@{}}1024 \\ (41\%)\end{tabular} & \multicolumn{1}{c|}{\textbf{\begin{tabular}[c]{@{}c@{}}45 \\ (3.6\%)\end{tabular}}}          & \textbf{\begin{tabular}[c]{@{}c@{}}78 \\ (3.1\%)\end{tabular}}          \\ \hline
BRAM     & \begin{tabular}[c]{@{}c@{}}419\\ (46\%)\end{tabular}   & \begin{tabular}[c]{@{}c@{}}786\\ (29\%)\end{tabular}                  & \begin{tabular}[c]{@{}c@{}}---\\ (---)\end{tabular}     & \begin{tabular}[c]{@{}c@{}}543 \\ (60\%)\end{tabular}              & \begin{tabular}[c]{@{}c@{}}114 \\ (12.5\%)\end{tabular}   & \multicolumn{1}{c|}{\textbf{\begin{tabular}[c]{@{}c@{}}142.5 \\ (98\%)\end{tabular}}}        & \textbf{\begin{tabular}[c]{@{}c@{}}691.5 \\ (76\%)\end{tabular}}        \\ \hline
URAM     & ---                                                     & \begin{tabular}[c]{@{}c@{}}0\\ (0\%)\end{tabular}                     & \begin{tabular}[c]{@{}c@{}}---\\ (---)\end{tabular}      & ---                                                                 & ---                                                      & \multicolumn{1}{c|}{\textbf{\begin{tabular}[c]{@{}c@{}}51 \\ (80\%)\end{tabular}}}            & \textbf{---}                                                             \\ \bottomrule
\end{tabular}
}
 
\end{table}

\subsubsection{RBMM Engine Design} \label{sec:RBMM}
The RBMM engine is designed and optimized for efficient matrix multiplication of \emph{various matrix sizes} in both MHA and FFN modules. Therefore we only need one single RBMM engine to handle all the matrix multiplications in the Transformer. This flexible design greatly optimize the area and power efficiency. The engine integrates activation functions and quantization operations directly, with behavior determined by the RBMM mode.

As illustrated in Fig. \ref{fig:Architecture_all}, for each invocation of the RBMM engine, it produces $N_{pe}$ RBVM results. The number of invocations is different in different mode configurations which is set by the COBRA controller. Each invocation of the engine is scheduled at an interval of one clock cycle, i.e., the initiation interval (II) of the pipeline is 1, which means the latency of RBMM Engine execution in each invocation is perfectly overlapped, and the modules inside RBMM are fully pipelined. The RBMM Engine accepts one $d$-bit row datapack from matrix $\bf A$ with its associated $1\times l$ DC INPUT vector, $N_{pe}$ $d$-bit column datapacks from matrix $\bf B$, and a quantization $1\times d$ $bias$ vector as inputs. Each RBMM PE processes a single row datapack and column datapack pair, computing their RBVM result and updating the DC RETURN which includes $h$ DC HEADs and a DC FULL after the ACC PATH. The DC RETURN will be treated as DC INPUT in the subsequent RBMM if (0, 1) binarization scheme is being used. 

Within the RBMM Engine shown in Fig. \ref{fig:RBMM PE}, each RBMM PE consists of $h$ HEAD PEs that compute RBVM results for $d_h$-bit head datapacks, and the final $d$-bit datapack RBVM result is obtained by accumulating these head-level computations as formulated in Eq. (\ref{ACC_HEAD_PE}). As for the popcount unit in HEAD PE, 64×3-bit ROM-based 6:3 compressors can be used to compute the population count of every 36 bits. 
 Initially, six compressors determine the number of ones in each group of six input bits. In the subsequent stage, the six 3-bit outputs from these compressors are combined using another round of compressors by feeding in the third bits, second bits, and first bits separately. Finally, a shift-and-add operation is performed to accumulate the results of the third, second, and first bits, yielding the total popcount. 
For FFN computations, since the hidden size is $FF_{size}=Rd$, the RBMM PE return value is accumulated with the previous output. 

As for bit width, the HEAD PE result requires $(\log_{2} d_h +1)$ bits, while the RBMM Engine output uses $\max (\log_{2} FF_{size} +1, h) = B_o$ bits, enabling efficient multi-head computation where binary multi-head attention results can be concatenated into a $k$-bit return value. What's more, the quantized binary RBVM output in this case set all $B_o$-bit to represent ``0" or ``1" and when it comes to the integer output, all bits are utilized to represent the integer. Finally, the address of the return output value is controlled by the Read/Write unit.

Our RBMM engine supports six operation modes as follows. We visualize the places of using each mode in the Transformer model in Fig. \ref{fig:Binary_Transformer}, and the hardware design in Fig. \ref{fig:Architecture_all} and Fig. \ref{fig:RBMM PE}.


\noindent
\textbf{M1. MHA $Q$/$K$/$V$ Computation} ($l \times d \times d$): 
The $Q$/$K$/$V$ matrix feeds to Matrix $\bf A$ and the corresponding weight matrix serves as Matrix $\bf B$. The HEAD PE activates the ACC PATH the threshold/data width value is set to $d_h$ and the RBMM PE produces quantized binary outputs where $output \geq bias$.

\noindent
\textbf{M2. MHA Attention Score} ($h$ $ l \times d_h \times l$):
The $Q$ serves as Matrix $\bf A$, while the $K$ serves as Matrix $\bf{B}$. The $h$ heads are processed simultaneously as they appear consecutively in the input datapacks. The HEAD PE enables the CONCAT PATH. Each head's result undergoes threshold comparison in SPS and attention mask application through iterative comparison of the current row and column indices. Note that the attention mask can support multiple mechanisms, including padding masks, causal masks, and others. This is achieved by checking whether the current loop index exceeds a value, which determines whether the attention mask is applied to the subsequent elements. The threshold/data width value is adjusted to $T + d_h$. The RBMM PE outputs the $h$-bit concatenated value and $h$ updated DC HEADs for each head.

\noindent
\textbf{M3. MHA Context Output}  ($h$ $ l \times l \times d_h$):
Matrix $\bf A$ is the attention score matrix of one head, and Matrix $\bf{B}$ is the transposed $V$ $l$-bit datapacks. The lower $l$ bits of the $d$ bit are valid. The threshold/data width value is modified to $l/h$, and the bias is added with $l\mod h$. The HEAD PE activates the ACC PATH. The RBMM PE adds the DC INPUT because the attention score matrix is (0, 1) value. The output is quantized with $bias$, and the $h$ head output addresses are arranged continuously to form an $l \times d$ matrix.

\noindent
\textbf{M4. MHA Linear Layer} ($l \times d \times d$):
This layer shares the configuration of mode $\textbf {M1}$ but produces integer output for the following LayerNorm operation instead of the quantized binary output value.

\noindent
\textbf{F1. FFN Linear Layer I ($l \times d \times FF_{size}$)} and \textbf{F2. FFN Linear Layer II ($l \times FF_{size} \times d$)}:
In the first FFN layer, the unsigned quantization $bias$ is applied to represent the ReLU and binarization operation as described in Eq. (\ref{alg:bias}). The DC FULL is updated to count the zeros after ReLU. The second FFN hidden layer involves the DC INPUT for summation and returns integer values accumulated with previous output values. The threshold/data width values are set to $d_{h}$. While these larger RBMMs seem to require a larger buffer of size $l \times FF_{size}$, the consecutive RBMMs are optimized as follows:\par
\begin{small}
\begin{align}
\mathbf{E} &= \text{ReLU}(\mathbf{X} \otimes \mathbf{Y}) \otimes \mathbf{Z} \nonumber\\
&= \text{ReLU}(\mathbf{X} \otimes [\mathbf{Y}_1,  \dots,\mathbf{Y}_r,\dots, \mathbf{Y}_R]) \otimes [\mathbf{Z}_1, \dots, \mathbf{Z}_r,\dots,\mathbf{Z}_R]^\top \nonumber\\
&= \text{ReLU}(\mathbf{X} \otimes  \mathbf{Y}_1 ) \otimes  \mathbf{Z}_1 + \dots + \text{ReLU}(\mathbf{X} \otimes  \mathbf{Y}_R) \otimes \mathbf{Z}_R,
\end{align}
\end{small}
where $\mathbf{X}\in \{(-1, 1)\}^{l\times d}$, $\mathbf{Y} \in \{(-1, 1)\}^{d \times FF_{size}}$, $\mathbf{Z} \in \{(-1, 1)\}^{FF_{size} \times d}$ and $\text{ReLU}$ denotes the ReLU operation with unsigned binarization. This optimization requires only $R$ iterations of $\text{ReLU}(\mathbf{X} \otimes \mathbf{Y}_r)$ and $\text{ReLU}(\mathbf{X} \otimes\mathbf{Y}_r) \otimes\mathbf{Z}_r$ operations, each involving an $l\times d \times d$ RBMM. In this way, only two buffers of size $l \times d$ are required. 
Moreover, for the corner case of resource-limited platforms like KV260 that cannot support two buffers, 
the hardware design adheres to the original computational flow and utilizes off-chip DDR memory to store the additional $R$ intermediate matrices.

\subsubsection{On-chip and Off-chip Memory} \label{sec:hw_arch_buffer}
Considering the inefficient off-chip memory access throughputs of edge platforms, on-chip memory management requires careful optimization to meet hardware resource constraints. The binary-format Matrix $\bf A$ and Matrix $\bf B$ are buffered entirely in BRAM or URAM due to their relatively small size. The DC INPUT vector and quantization $bias$ vector are also maintained in on-chip memory. In contrast, the weight and bias for different RBMM and transformer layers are fetched from off-chip memory and loaded to Matrix $\bf B$ and $bias$ vector. 
However, the output matrix necessitates integer representation to support LayerNorm operations. Thus, a minimum of three $B_o$-bit matrices with dimensions $l \times d$ are required to store the $Q$/$K$/$V$ projection output matrices and support the two intermediate matrices in mode \textbf{F1} and \textbf{F2}. 
These matrices are allocated to off-chip DDR memory or on-chip BRAM/URAM based on available resources. 
For instance, with parameters $l=512$ and $d=768$ using $B_o = 13$-bit integer representation, the KV260 platform can only accommodate one output matrix in on-chip memory, necessitating off-chip DDR allocation for the remaining matrices. This configuration leads to performance degradation due to the inefficient DDR read/write operations through the AXI master interface on edge platforms.
In contrast, the ZCU102, with its larger memory capacity, can buffer all three intermediate output matrices in on-chip memory, enabling optimal performance.



\begin{table}[t]
 \centering
 \caption{Resource Breakdown for COBRA ($N_{pe}=32$). }
\label{tab:resource_breakdown}
    \resizebox{\linewidth}{!}{
\begin{tabular}{c|cccc}
\toprule
COBRA $N_{pe}=32$                                                                 & LUT                                                       & FF                                                        & DSP                                                  & BRAM                                                      \\ \hline
RBMM Engine                                                                        & \begin{tabular}[c]{@{}c@{}}42926\\ (34.96\%)\end{tabular} & \begin{tabular}[c]{@{}c@{}}62782\\ (38.66\%)\end{tabular} & 0                                                    & 0                                                         \\ \hline
LayerNorm Unit                                                                     & \begin{tabular}[c]{@{}c@{}}14055\\ (11.45\%)\end{tabular} & \begin{tabular}[c]{@{}c@{}}13505\\ (8.32\%)\end{tabular}  & \begin{tabular}[c]{@{}c@{}}78\\ (100\%)\end{tabular} & \begin{tabular}[c]{@{}c@{}}16\\ (2.31\%)\end{tabular}     \\ \hline
Data Packing Conversion Units                                                           & \begin{tabular}[c]{@{}c@{}}27187\\ (22.14\%)\end{tabular} & \begin{tabular}[c]{@{}c@{}}52707\\ (32.46\%)\end{tabular} & 0                                                    & 0                                                         \\ \hline
\begin{tabular}[c]{@{}c@{}}On-chip Memory, \\ AXI Bus and COBRA CTRL\end{tabular} & \begin{tabular}[c]{@{}c@{}}38634\\ (31.46\%)\end{tabular} & \begin{tabular}[c]{@{}c@{}}33399\\ (20.57\%)\end{tabular} & 0                                                    & \begin{tabular}[c]{@{}c@{}}675.5\\ (97.67\%)\end{tabular} \\ \hline
Total                                                                              & 122802                                                    & 162393                                                    & 78                                                   & 691.5                                                     \\ \bottomrule
\end{tabular}}
\end{table}

\section{Experimental Results and Analysis}
\subsection{Experimental Setup}
We implement COBRA accelerator using high-level synthesis C/C++ in Vitis HLS  and Vivado 2023.1. We evaluate our design on AMD KV260 (Zynq UltraScale+ XCK26 MPSoC) and AMDZCU102 (Zynq UltraScale+ XCZU9EG MPSoC) boards. 
For Vitis HLS synthesis, we set the target frequency to 333MHz, achieving a maximum of 389 MHz on the KV260 and 339 MHz on the ZCU102. To ensure timing closure, we use 300 MHz for the final bitstream generation.
We evaluate COBRA with BERT-base model: the maximum sequence length $l=512$, the hidden dimension $d=768$, the number of head $h=12$, the FFN hidden dimension $FF_{size}=4*d = 3072$ and 12 transformer layers in total. 

\subsection{COBRA Accelerator Evaluation and Comparison}
\begin{figure}[t]
    \centering
    \includegraphics[width=0.9\linewidth]{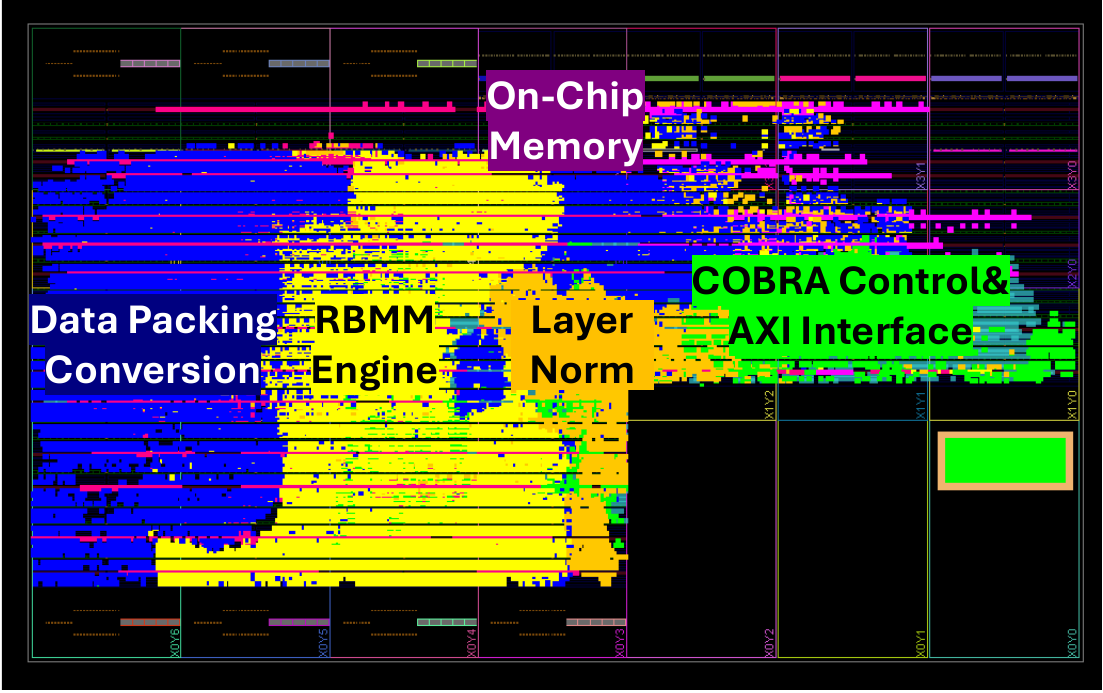}
    \caption{COBRA Accelerator Implementation Layout (ZCU102)}
    \label{fig:floorplan}
\end{figure}
As shown in Table \ref{tab:performance}, our FPGA implementation of COBRA achieves energy efficiency improvements of 150.7× and 144.4× on the KV260 platform compared to the BiT \cite{liu2022bit} and COBRA GPU baselines 
 (PyTorch GPU), respectively. On the ZCU102 platform, it achieves energy efficiency improvements of 325.1× and 311.6×, respectively. 
The resource utilization shown in Table \ref{tab:resource utilization} reveals that the configuration with $N_{pe} = 16$ on KV260 consumes the least LUT resources while maintaining energy efficiency comparable to BETA \cite{ji2024beta} and BAT \cite{ji2024co} implementations. These demonstrate the adaptability and efficiency of our hardware architecture on resource-constrained edge devices.
The $N_{pe} = 32$ COBRA implementation on mid-end ZCU102 delivers the best performance in both energy efficiency and throughput among all evaluated implementations. Specifically, it achieves 3.5× throughput and 2.2× energy efficiency improvements, compared to BAT \cite{ji2024co}, the state-of-the-art binary transformer accelerator. 
Furthermore, our design on the low-power ZCU102 platform achieves  $2.5\times$ higher throughput than COSA \cite{COSA} and TransFRU \cite{TransFRU} designs on high-performance datacenter-grade  FPGAs, while our design consumes significantly lower power and has reduced LUT or DSP resource utilization compared to theirs. 
The resource breakdown of COBRA $N_{pe}=32$ is shown in Table \ref{tab:resource_breakdown}. Our RBMM Engine design uses 42,926 LUTs for efficient quantization-fused RBMM. The LayerNorm unit occupies 78 DSPs for computing the standard deviation and normalization. Notably, the data packing conversion units consume significant resources due to the LUT-based intermediate buffers in the transpose operation for 
$V$ output matrix to datapacks (details in the \textbf{M3. MHA Context Output} section).
Fig. \ref{fig:floorplan} shows the implementation layout of the COBRA accelerator on the ZCU102 FPGA. 
Major components are highlighted and labeled in the figure.

\par
\begin{table}[t]
 \centering
\caption{Impact of Each Proposed Optimization.}
\label{tab:ablation}
    \resizebox{\linewidth}{!}{
\begin{tabular}{c|c|c|c|cc}
\toprule
\multirow{2}{*}{Design}                                                     & \multirow{2}{*}{\begin{tabular}[c]{@{}c@{}}Throughput \\ (GOPS)\end{tabular}} & \multirow{2}{*}{\begin{tabular}[c]{@{}c@{}}Power\\ (W)\end{tabular}} & \multirow{2}{*}{\begin{tabular}[c]{@{}c@{}}Energy Efficiency \\ (GOPS/W)\end{tabular}} & \multicolumn{2}{c}{Hardware Resource}                                                                                               \\ \cline{5-6} 
                                                                            &                                                                               &                                                                      &                                                                                        & \multicolumn{1}{c|}{LUT}                                                   & DSP                                                    \\ \hline
\textbf{\begin{tabular}[c]{@{}c@{}}COBRA \\ $N_{pe}=32$\end{tabular}}      & \textbf{3894.74}                                                              & \textbf{8.68}                                                        & \textbf{448.70}                                                                        & \multicolumn{1}{c|}{\textbf{122K}}                                         & \textbf{78}                                            \\ \hline
w/o SPS                                                                     & \begin{tabular}[c]{@{}c@{}}6.91\\ (564$\times$)\end{tabular}                          & \begin{tabular}[c]{@{}c@{}}13.41\\ (+4.7)\end{tabular}               & \begin{tabular}[c]{@{}c@{}}0.52\\ (862.9$\times$)\end{tabular}                                & \multicolumn{1}{c|}{\begin{tabular}[c]{@{}c@{}}158K\\ (+36K)\end{tabular}} & \begin{tabular}[c]{@{}c@{}}104\\ (+26)\end{tabular}    \\ \hline
\begin{tabular}[c]{@{}c@{}}w/o 6:3 compressor\\ based popcount\end{tabular} & \begin{tabular}[c]{@{}c@{}}3873.91\\ (1.005$\times$)\end{tabular}                    & \begin{tabular}[c]{@{}c@{}}12.18\\ (+3.5)\end{tabular}               & \begin{tabular}[c]{@{}c@{}}318.06\\ (1.4$\times$)\end{tabular}                                & \multicolumn{1}{c|}{\begin{tabular}[c]{@{}c@{}}134k\\ (+12K)\end{tabular}} & \begin{tabular}[c]{@{}c@{}}1614\\ (+1536)\end{tabular} \\ \hline
\begin{tabular}[c]{@{}c@{}}w/o pipeline \\ for RBMM Engine\end{tabular}     & \begin{tabular}[c]{@{}c@{}}794.32\\ (4.9$\times$)\end{tabular}                       & \begin{tabular}[c]{@{}c@{}}7.98\\ (-0.7)\end{tabular}                & \begin{tabular}[c]{@{}c@{}}99.54\\ (4.5$\times$)\end{tabular}                                 & \multicolumn{1}{c|}{\begin{tabular}[c]{@{}c@{}}125K\\ (+3K)\end{tabular}}  & 78                                                     \\ \bottomrule
\end{tabular}}
\end{table}
\subsection{Ablation Study of COBRA Hardware Design}\label{subsec:ablation}
We performed an ablation study on our proposed optimization techniques to better understand their impact on the overall accelerator performance. Table \ref{tab:ablation} shows the results. 
\subsubsection{SPS instead of Softmax}
As discussed in Section \ref{sec:sps}, the Softmax hardware module is inefficient for hardware. In contrast, our proposed SPS-fused RBMM achieves a throughput improvement of 564$\times$ compared to the original Softmax unit, which requires additional 36K LUTs and 26 DSPs for the computation of attention scores across \( h \) heads after RBMM.

\subsubsection{6:3 Compressor-Based Popcount instead of CPU-Optimized Popcount}
The design of the Popcount unit is a critical consideration for the RBMM Engine, as it is replicated \( h \) times in each RBMM PE. The CPU-optimized popcount, as introduced in \cite{hamming_weight_wikipedia}, leads to additional 12K LUTs and 1536 DSPs due to the inclusion of multipliers.

\subsubsection{Pipeline Execution of RBMM Engine instead of Serial Execution}
As outlined in Section \ref{sec:RBMM}, the components within the RBMM Engine are fully pipelined, and the engine's initiation interval (II) is one clock cycle. If pipeline scheduling is disabled, the RBMM Engine's execution after each call is serialized, resulting in only 20\% of the throughput achieved by the pipelined version. 

\section{Conclusion}
In this paper, we propose \textbf{COBRA}, a co-designed Binary Transformer Accelerator optimized for edge FPGAs. Our approach incorporates a hardware-friendly Shifted Polarized Softmax (SPS) design, a Real 1-bit Binary Matrix Multiplication Engine (RBMM), and an adaptive computational flow, enabling us to fully harness the potential of binary transformer acceleration. COBRA achieves a 3.5× improvement in throughput and a 2.2× gain in energy efficiency with negligible accuracy degradation compared to state-of-the-art binary transformer accelerators. These on-board evaluation results highlight the performance of our design, enabling highly efficient  edge inference of transformer models. 
\bibliography{refernces.bib}{}

\bibliographystyle{IEEEtran}

\end{document}